# Probing multiphoton light-induced molecular potentials

M. Kübel,[1,2,3], M. Spanner[1], Z. Dube[1], A.Yu. Naumov[1], S. Chelkowski[4], A.D. Bandrauk[4],

M.J.J. Vrakking[5], P.B. Corkum[1], D.M. Villeneuve[1], and A. Staudte[1]

[1] Joint Attosecond Science Laboratory, National Research Council and University of Ottawa, 100 Sussex Drive, Ottawa, Ontario K1A 0R6, Canada

[2] Department of Physics, Ludwig-Maximilians-Universität Munich, Am Coulombwall 1, D-85748 Garching, Germany

[3] Institute for Optics and Quantum Electronics, University of Jena, D-07743 Jena, Germany

[4] Laboratoire de Chimie Théoretique, Faculté des Sciences, Université de Sherbrooke, Sherbrooke, Québec J1K 2R1, Canada

[5] Max-Born-Institute, Max-Born-Straße 2A, D-12489 Berlin, Germany

**Abstract.** The strong coupling between intense laser fields and valence electrons in molecules causes a distortion of the potential energy hypersurfaces which determine the motion of nuclei in a molecule and influences possible reaction pathways. The coupling strength varies with the angle between the light electric field and valence orbital, and thereby adds another dimension to the effective molecular potential energy surface, allowing for the emergence of light-induced conical intersections. Here, we demonstrate in theory and experiment that the full complexity of such light-induced potential energy surfaces can be uncovered. In $H_2^+$, the simplest of molecules, we observe a strongly modulated angular distribution of protons which has escaped prior observation. These modulations directly result from ultrafast dynamics on the light-induced molecular potentials and can be modified by varying the amplitude, duration and phase of the mid-infrared dressing field. This opens new opportunities for manipulating the dissociation of small molecules using strong laser fields.

## Introduction

Potential energy surfaces describe the forces acting on the nuclei of a molecule. Within the Born-Oppenheimer approximation the motion of the nuclei along these potentials is treated independently of the electronic motion. This picture breaks down when the electronic level separation becomes comparable to the kinetic energy of the nuclei. This occurs at specific points in the molecular geometry, which are known as conical intersections and which are a hallmark of polyatomic molecules [1]. Conical intersections play an eminent role in visible and ultraviolet photochemistry [2,3], for example in isomerization [4,5], and electron transfer processes [6]. Moreover, they are strongly implicated in the photostability of DNA by way of allowing radiation-less de-excitation [7].

The single-photon transition between two dipole-coupled electronic states can also create a conical, albeit transient, intersection. Hence, these localized features of the laser-dressed potential energy surface have been dubbed light-induced conical intersections (LICI) [8,9]. Their precise position and the underlying dipole coupling strength are determined by the frequency and intensity of the incident light. LICIs can also be found in diatomic molecules, since the angle between the light polarization and the molecular axis adds a another degree of freedom to the nuclear motion, [10,11]. The angle-dependent distortion of the molecular potential energy surfaces in a linearly polarized laser field directly affects molecular dissociation [12–16] and has been predicted to cause rotational excitation [17–20]. Recently, experimental indications of LICI in $H_2^+$ have been found in angle-resolved ion spectra [21].

In ultrashort infrared laser fields, the light intensity can easily exceed the threshold for multiphoton transitions. While LICIs are a consequence of single-photon couplings and therefore the potential energy scales linearly with respect to variations of the laser field strength, multiphoton couplings lead to unique structures of their own. In the case of diatomic molecules these structures become non-linear point intersections of the potential energy surfaces. The one-dimensional treatment of single and multiphoton resonances has led to the prediction of light-induced potentials (LIPs) [21–27]. However, the consequences of the angle-dependent coupling strength around non-linear point intersections for the dissociation dynamics have so far been largely unexplored.

Here, we show in theory and experiment that non-linear light-induced point intersections can result in strong modulations of the angular ion yield. We choose the simplest molecule, $H_2^+$, which is widely regarded as a prototype system for the interaction of molecules with light [28]. Due to the sparsity of electronic states, $H_2^+$ can often be described as a two-level system consisting of the two lowest electronic states, *1sσ$_g$* and *2pσ$_u$*. When coupled to intense laser fields these states give rise to intensity-dependent dissociation mechanisms known as bond softening [29] and above threshold dissociation [30]. The possibility to control them using the laser amplitude, frequency, phase and pulse duration has been demonstrated [31–35]. In particular, the opposite parity of the $\sigma_g$ and $\sigma_u$ states can lead to electron localization [36], giving rise to charge resonance-enhanced ionization [37,38] and symmetry breaking in dissociation [39–41] under the influence of two-color laser fields [42,43] or carrier-envelope phase stable few-cycle pulses [44–47]. Descriptive treatments for most of these phenomena are provided by dressed-state pictures, such as LIPs.

## Results and Discussion

### Non-linear point intersections

Figure 1(a) shows an example of some LIP energy surfaces for $H_2^+$ in moderately intense ($3 \cdot 10^{13}$ W/cm$^2$), visible (685 nm) laser field, calculated using the procedure outlined in the Methods section. Shown are the laser-dressed states $\sigma_g$ and $\sigma_u$-*1ħω*. Along the laser polarization, i.e., *θ = 0, π,* the state crossing indicated in Figure 1(b) opens up and turns into an avoided crossing. This necessarily lowers the potential barrier at the avoided crossing permits and permits dissociation of formerly bound molecules (bond softening). Importantly, no such avoided crossing occurs when the laser polarization is perpendicular to the internuclear axis. Therefore a LICI is formed at the internuclear distance where the laser-dressed $\sigma_g$ and $\sigma_u$-1 states cross.

While single-photon transitions dominate in moderately intense visible laser fields, multiphoton transitions become relevant already at moderate intensities, when the wavelength is shifted into the mid-infrared [48]. For example, the three-photon transition by 2300 nm light becomes significant already at an intensity of approximately $5 \cdot 10^{12}$ W/cm$^2$, see Supplementary Figure 1. Several crossings of potential curves that correspond to multiphoton transitions between the $\sigma_g$ and $\sigma_u$ states of H$_2^+$ at a wavelength of 2300 nm are shown in Figure 1(c). The corresponding LIP energy landscape calculated for an intensity of $3 \cdot 10^{13}$ W/cm$^2$ is presented in Figure 1(d). The potential energy surfaces exhibit complex structures that result from multi-photon couplings. Indicated in the figure are the curve crossings due to n-photon ($n$ = 1, 3 and 5) transitions. Notably, the shapes for n =3, 5 deviate significantly from the conical shapes of LICIs (n = 1), as a direct consequence of the associated *n*-photon non-linearities. In order to see how these light-induced structures affect the nuclear dynamics, we first solve the 2-dimensional time-dependent Schrödinger equation (2D TDSE, see Methods for details) for H$_2^+$ under the influence of a moderately intense mid-IR laser pulse. The calculated proton momentum distributions presented in Figure 1(e) show distinct features in the proton angular distribution that can be associated with the *n*-photon couplings. While these features are absent in the single-photon coupling regime, significantly higher intensities produce very convoluted dissociation patterns that would likely defy experimental resolution, see Supplementary Figure 2.

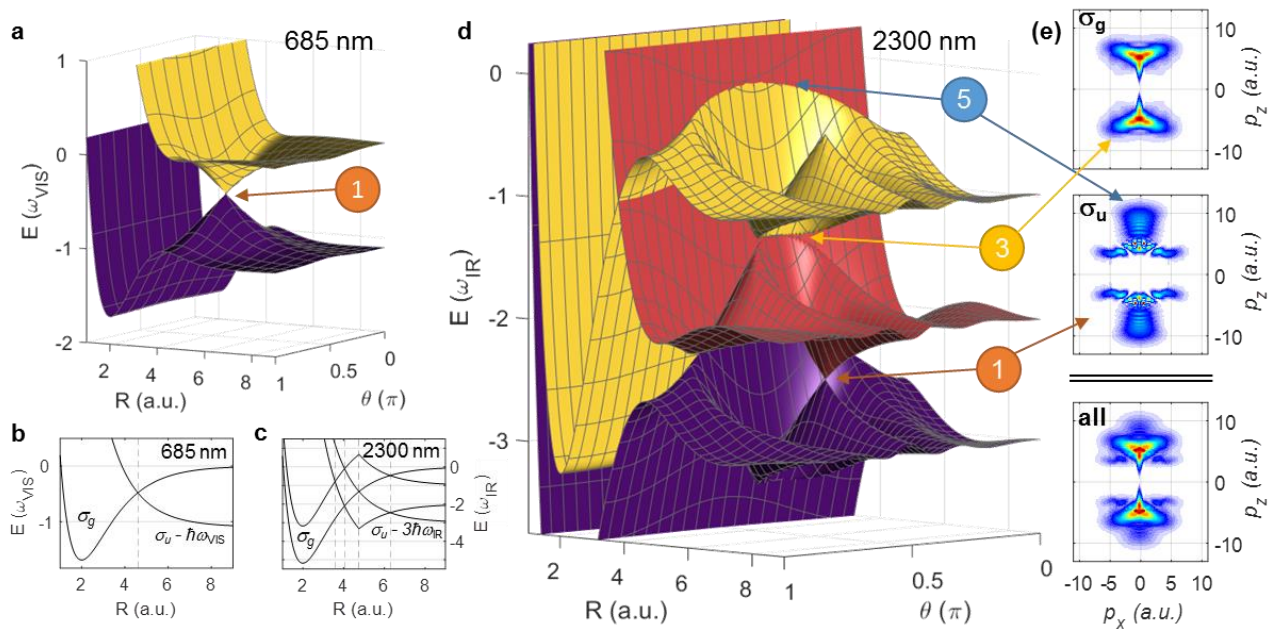

**Figure 1.** Diabatic single- and multiphoton light-induced molecular potentials. (a) Angle-dependent potential energy surfaces of $H_2^+$ dressed by a linearly polarized, moderately intense, visible (685 nm, $3\times10^{13}$ Wcm$^{-2}$) laser field. The LICI at the one-photon crossing is marked by a 1. The potential energy curve shown in (b) corresponds to a lineout along the internuclear distance R through the LICI. The dotted line marks the position of the one-photon resonance between the $\sigma_g$ and $\sigma_u$ states. The potential energy diagram in (c), shows several multiphoton $\sigma_g$ (dissociation limits of (0, -2, -4,...) $\omega_{IR}$) and $\sigma_u$ (dissociation limits of (1, -1, -3, ...) $\omega_{IR}$) states in a 2300 nm dressing field. Two of these states are labeled, and he location of multi-photon transitions of order 7, 5, 3, and 1 (from left to right) are indicated. (d) Same as (a) for a mid-IR dressing field (2300 nm, $3\times10^{13}$ Wcm$^{-2}$). Structures attributed to 1, 3, and 5 photon couplings are marked. (e) Calculated proton momentum distributions produced by a 2300 nm, $3\times10^{13}$ Wcm$^{-2}$ laser field. Contributions from dissociation on the $\sigma_g$ and $\sigma_u$ surfaces are shown separately

## *Structured proton angular distribution*

In order to experimentally probe the light-induced molecular potentials depicted in Fig. 1(d,e), describing the situation where the mid-IR field induces multiphoton dynamics in the dissociation process, but does not cause ionization, we implement the two-pulse scheme depicted in Fig. 2(a). First, an intense, few-cycle visible laser pulse ionizes neutral $H_2$, producing a bound coherent wave packet in $H_2^+$ with a nearly isotropic alignment with respect to the laser polarization [49]. Second, a moderately intense mid-IR pulse creates the LIPs on which dissociation occurs. The LIPs are probed by recording the momentum distribution of protons resulting from the dissociating part of the molecular wave packet. The molecular ions dissociate along their initial alignment direction, unless rotational dynamics occur as predicted e.g. in Refs [17–20].

The intent of the two-pulse scheme is to decouple the production of the molecular wavepacket from the field that generates the LIPs. Thus, scanning the time delay between the laser pulses allows for probing the LIPs at selected times within the mid-IR pulse. Moreover, the use of a shorter wavelength pulse for ionization allows us to reduce the focal volume averaging in the long-wavelength field, which often washes out subtle features in strong-field experiments (e.g., [50]). Finally, choosing a perpendicular relative polarization of the visible and mid-IR pulses is expected to avoid overlap between the signal of interest produced by the mid-IR pulse, and any protons produced by the visible pulse alone. The experimental set-up is described in the Methods section *Experiment*.

Figure 2(b) shows experimental results obtained with the cross-polarized few-cycle visible and mid-IR pulses. The 3D momentum distributions of protons and electrons were measured in coincidence using Cold Target Recoil Ion Momentum Spectroscopy (COLTRIMS). The coincidence measurement allows us to present results in the recoil frame, where the electron recoil has been removed from the measured ion momentum distribution. The results exhibit a striking angular structure which is blurred if the recoil momentum is not accounted for (see Supplementary Figure 5). The angular structure consists of on-axis features along either of the polarization axes, and additional spots at intermediate angles. Drawing a comparison to results obtained with only visible (bottom inset in Fig. 2(b)) or (more intense, $1 \cdot 10^{14}$ W/cm$^2$) mid-IR (top inset in Fig. 2(b)) pulses, suggests that the on-axis features arise from bond softening $H_2^+$ by either the visible or mid-IR pulses alone. Note, that the signal along the mid-IR polarization in the two-color experiment does not arise from dissociative ionization of neutral $H_2$ by the mid-IR pulse on its own, as no notable ionization of neutral $H_2$ is obtained at the intensity of $3 \cdot 10^{13}$ W/cm$^2$. Hence, the comparative mid-IR only data (top inset in Fig. 2(b)) is presented for a higher intensity of $1 \cdot 10^{14}$ W/cm$^2$. The additional spots in the two-color data are tentatively attributed to dynamics caused by the light-induced structures in the molecular potential energy landscape, cf Fig. 1(d).

Surprisingly, the experimental results presented in Figure 2(b) exhibit a much more pronounced angular structure than the TDSE results for the mid-IR field alone, presented in Figure 1(e). A hint on the origin of the additional angular structure in the experiment comes from measurements carried out with a parallel polarization of the visible and mid-IR pulses. The proton momentum distribution obtained with parallel polarization is shown in Figure 2(c). It resembles the results obtained with mid-IR pulses only and does not exhibit significant angular structure. Although the intention of our scheme was to decouple the effects of the mid-IR and VIS pulses, the striking dependence of the angular structure on the relative polarization of the two laser fields implies that the visible field contributes to the formation of the additional spots and must be considered in further investigations.

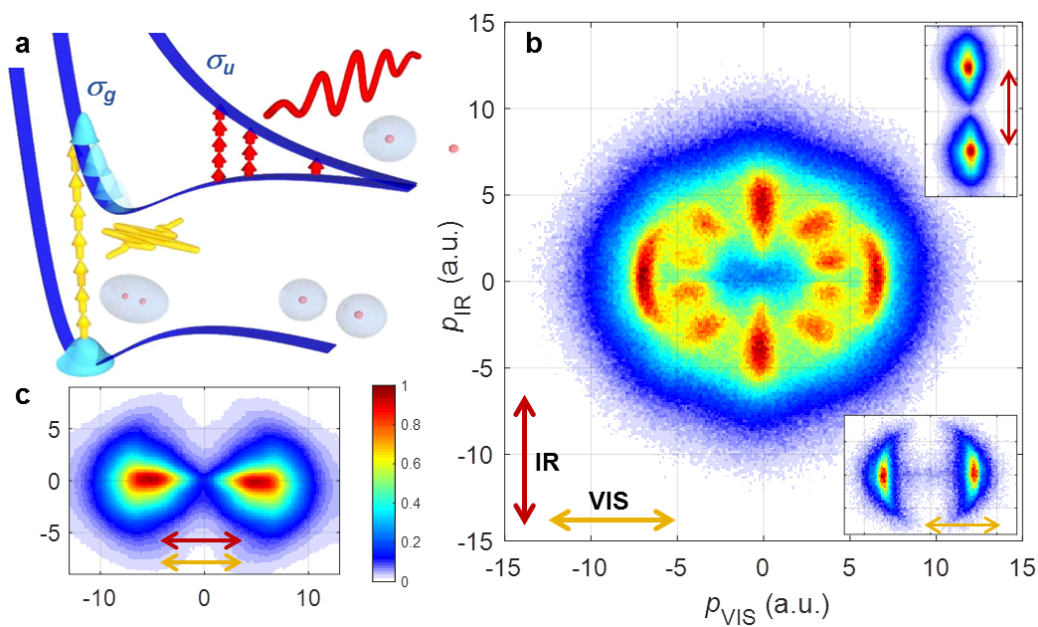

Figure 2. Experimental signatures of light-induced potentials. (a) A few-cycle visible pulse (yellow) ionizes $H_2$ and prepares a wave packet on the $\sigma_g$ state of the molecular cation. An additional mid-infrared control pulse (red) couples the $\sigma_g$ and $\sigma_u$ states by 1, 3, or 5 photons, creating light-induced potentials, on which $H_2^+$ dissociates into an H atom and a proton. (b) The measured proton momentum distribution in the recoil frame for perpendicularly polarized two-color fields (730 nm, 5 fs, $2\times10^{14}$ Wcm$^{-2}$, and 2300 nm, 65 fs, $3\times10^{13}$ Wcm$^{-2}$) probes the light-induced potentials. It strongly contrasts results obtained for only mid-IR pulses (top, 2000 nm, 65 fs, $1\times10^{14}$ Wcm$^{-2}$) or only visible bottom, 730 nm, 5 fs, $2\times10^{14}$ Wcm$^{-2}$) pulses. The angular structure is moreover absent in two-color experiments carried out with parallel polarization (c). The data has been integrated over the direction perpendicular to the polarization plane. The arrows indicate the polarization axes of the visible (orange) and mid-IR (red) laser pulses. The color map represents the measured proton yield normalized to the maximum value in each plot.

## Numerical results

In a first step to understand the dynamics producing the structured proton angular distribution observed in cross-polarized fields, we solve the 2D TDSE for $H_2^+$, taking both laser pulses into account, see Methods for details. For the visible few-cycle pulse, we also consider a weak pulse pedestal at 5% of the peak intensity. The initial alignment of the molecular axis with respect to the laser polarization is assumed to be istotropic.

It has been recognized in the literature that angular modulations in the proton spectra can arise from rotational dynamics in the vicinity of the LICI [18–21]; more specifically, simulations that include rotational motion in the dissociation dynamics show angular modulations that are absent when the rotational degrees of freedom are frozen. These modulations can be connected to rotational scattering of the dissociating wavepacket from the LICI [18–21]. A first candidate for the physical mechanism underlying the appearance of a structured proton angular distribution is therefore the formation of a high-order rotational wavepacket in the dissociating molecular cation. In order to test the role of rotational dynamics, we perform a first set of calculations where rotational transitions are artificially switched off and present the results in Fig. 3(a). Evidently, strong modulations in the angular distribution are obtained, even without the inclusion of wavepacket rotation. This suggests that rotational dynamics are not the primary physical mechanism underlying the angular modulation in the proton momentum distribution. Therefore, it will be important to identify how angular modulations arise already within a 1D treatment.

The second set of calculations (Figure 3(b)) takes rotational dynamics into account. The differences between Figures 3(b) and 3(a) show the impact of rotations in certain parts of the momentum distributions. The black arrows highlight pronounced differences between the results of the two calculations at angles $\theta = 0°$, and $\theta = 20°$. These differences illustrate the role of rotational alignment in shaping the final momentum distribution. On the basis of the comparison between Figure 3a and 3b, we conclude that rotational dynamics play a significant but secondary role in defining the final momentum distribution; in contrast with the previously considered pure LICI case, the addition of rotations is not the sole cause of the angular structures in our experiment.

In order to identify the essential mechanism creating the angular structure in the absence of rotations, we employ two-color Floquet theory (see Methods). We calculate the angle-dependent field-dressed states of $H_2^+$ using a two-color laser field, $\vec{E}(t) = \sqrt{I_{VIS}}\cos(\omega_{VIS}t + \varphi_{VIS})\vec{x} + \sqrt{I_{IR}}\cos(\omega_{IR}t + \varphi_{IR})\vec{y}$.

The field consists of a moderately intense mid-infrared field ($\lambda_{IR}$ = 2280 nm, $I_{IR}$ = 3×10$^{13}$ Wcm$^{-2}$) and a weak visible field ($I_{VIS}$ = 1×10$^{13}$ Wcm$^{-2}$), corresponding to the pulse pedestal used in the TDSE calculations. We take $\lambda_{VIS} = \lambda_{IR}$ / 3 to ensure the periodicity of the laser fields required by Floquet theory. The resulting light-induced potential energy landscape depends on the relative optical phase, $\Delta\varphi = \varphi_{VIS} - \varphi_{IR}$. Exemplarily, we present the light-induced potential energy landscape for $\Delta\varphi = 0$ in Fig. 3(c).

A detailed analysis and discussion of the results from Floquet theory is presented in Supplementary Note 4. In brief, we find that the Floquet states represent a conclusive basis for understanding the emergence of the angular structure in the proton momentum distribution, even without rotational dynamics taken into account. The following picture can be invoked.

As the alignment angle of the molecular axis in the polarization plane (see Fig. 3), $\theta$, is increased, the field components parallel to the molecular axis vary as $E_{VIS}(\theta) = E_{0,vis}\cos(\theta)$, and $E_{IR}(\theta) = E_{0,IR}\sin(\theta)$. This leads to a pronounced angle dependence of the field-dressed potential energy curves (see Figure 3c), where several Floquet state crossings open up and close again as $\theta$ is varied. Specifically, at $\theta$ = 0°, i.e. for alignment of the molecular axis perpendicular to the mid-IR polarization, the effect of the mid-IR field is insignificant and dissociation proceeds as in the single-color case (seen in Fig. 2(b)). Specifically, the wavepacket dissociates on the purple surface in Fig. 3(c). As $\theta$ is increased, a new dissociation channel due to one-photon coupling by the mid-IR field opens up. The new channel competes with the original one, which moves population to the pronounced feature at $\theta = 10°$, making the on-axis feature much narrower than in the single-color case. This new dissociation channel corresponds to dissociation on the red surface in Fig. 3(c). As $\theta$ is further increased, the width of the avoided crossing reaches $2\omega_{IR}$, which closes the dissociation channel and gives rise to a LICI at $\theta \sim 30°$, clearly visible in Fig. 3(c).

Notably, the computational results obtained without (Fig. 3(a)) and with (Fig. 3(b)) rotations strongly differ at $\theta \sim 20°$. We attribute this to the presence of the LICI which promotes strong rotational dynamics, as the nuclear wave packet propagates around the cone in the LIP landscape. In a similar manner, the splitting of the narrow feature at 0° in Fig. 3(a) into the double peak structure in Fig. 3(b) is attributed to the point intersections at $\theta = 0°$.

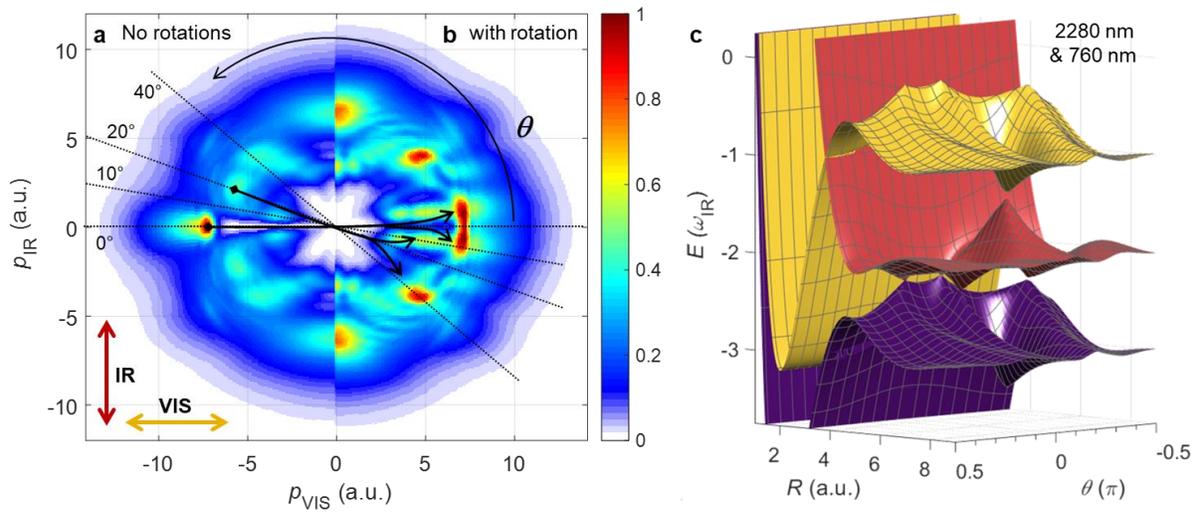

**Figure 3 Numerical results for two-color bond-softening of $H_2^+$.** Proton momentum distributions obtained from solving the time-dependent Schrödinger Equation (a) without and (b) with rotational dynamics taken into account. The black arrows indicate population transfer through rotational dynamics. The dotted black lines at various angles are drawn as a guide to the eye. (c) The light-induced potential energy landscape obtained from Floquet theory for a two-color field with relative phase $\Delta\varphi = 0$.

## *Delay dependence*

Scanning the time delay between the visible and mid-IR pulses in our experiment allows us to probe the variations in the LIPs throughout the mid-IR pulse. The time delay controls the time of ionization, which determines the (i) strength, (ii) duration, and (iii) phase of the mid-IR field at the time it interacts with the molecular ion. In Figure 4, we analyze the fragment momentum distribution for overlap of the ionizing visible pulse with the rising edge, the maximum and the falling edge of the mid-IR laser pulse. Each of the presented spectra are integrated over two mid-IR cycles and therefore not expected to be sensitive to the mid-IR phase.

Figure 4(a) shows the vector potential of the mid-IR pulse used in our experiment, as measured with the STIER technique [51] (see Supplementary Figure 4). Selected recoil-frame proton momentum distributions are presented in Figs 4 (b-d). The delay-dependent results probe the evolution of the LIP energy landscape throughout the mid-IR pulse. This is evidenced by the changes in the recorded dissociation patterns, as the delay between visible and mid-IR pulses is varied. For example, the feature at $\theta = 90°$, i.e., along the mid-IR polarization axis, peaks around the center of the pulse, Fig. 4(c), where it represents the dominant contribution to the proton momentum distribution. When the ionization occurs on the falling edge of the mid-IR pulse (Figure 4(d)), the 90° feature is absent. On the basis of the computational results presented in Fig. 1 and the two-color Floquet states shown in Figure 3(c), we attribute this peak to a five-photon coupling induced by the mid-IR pulse. The nonlinearity of this process explains why this feature is particularly visible near the maximum of the mid-IR intensity. Notably, the maximum yield of protons emitted at 90° is obtained when the visible pulse precedes the peak of the mid-IR pulse by (8.3 ± 0.5) fs, in reasonable agreement with the 7.3 fs vibrational half-period of $H_2^+$ [52]. For earlier delays, when ionization occurs on the rising edge of the mid-IR pulse (Figure 4(b)), the weaker signal at 90° indicates that dissociation occurs before the molecular ion interacts with the center of the mid-IR pulse. Similar observations are made for the feature at intermediate angles (around $\theta\sim40°$ in Fig. 3(a)). Moreover, its angular position also varies from $\theta\sim30°$ in Fig. 4(b) towards $\theta\sim40°$ in Fig. 4(c).

Contrary to the non-linear features, the feature at $\theta=10°$, i.e. close to the visible polarization axis, exhibits little delay dependence. As discussed above, this feature can be understood as a consequence of the single-photon couplings by both, the visible and the mid-IR fields. The absence of non-linearity in this process explains the insensitivity of the 10° feature to the mid-IR intensity.

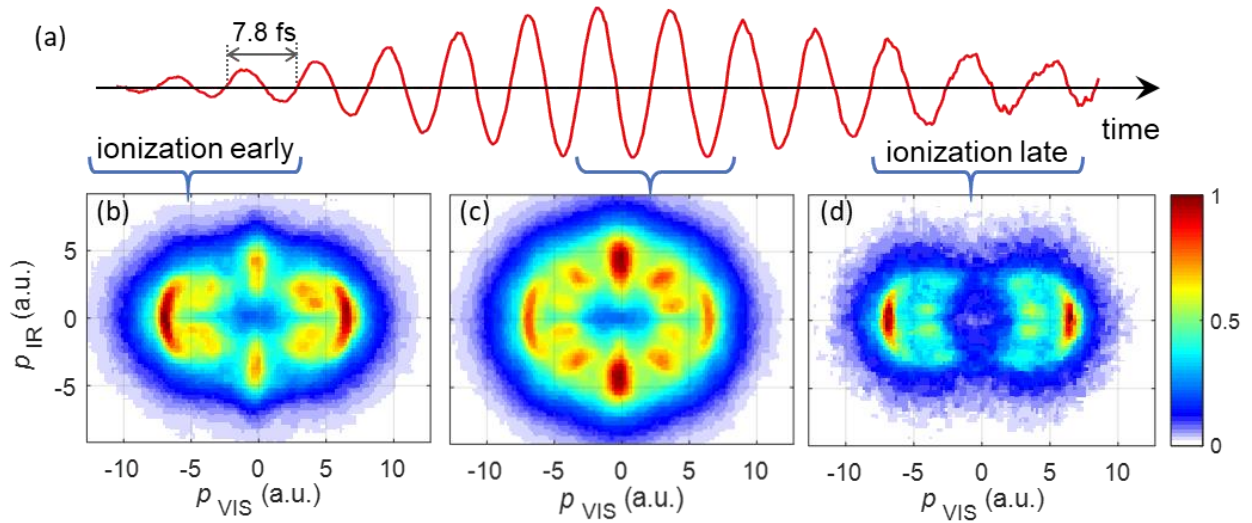

Figure 4. Tracking the evolution of light-induced molecular potentials in $H_2^+$ throughout a mid-IR laser pulse. (a) Vector potential of the mid-IR dressing field measured using the STIER technique. (b-d) Measured $H^+$ momentum distributions in the polarization plane for different ionization times within the mid-IR pulse. The signal has been integrated over the delay ranges indicated by the brackets. The colorbar indicates the proton yield normalized to the maximum in each plot.

## Summary and Outlook

In summary, we have demonstrated a powerful approach for probing light-induced molecular potentials. We observed strongly modulated proton angular distributions in experiments were $H_2^+$ ions produced by a linearly polarized, few-cycle, visible laser pulse, are dissociated by a cross-polarized mid-IR laser field. We have shown that the modulations can be understood as signatures of complex light-induced potential energy landscapes that are shaped by both single-photon and multiphoton transitions. Specifically, the modulations arise from a combination of two effects: First, dissociation pathways for a given mid-IR laser intensity open and close as a function of alignment angle; second, rotational motion around light induced point intersections, such as LICIs, shape the modulated angular ion yield.

Probing the LIPs produced by the mid-IR dressing field on its own may be improved by using a shorter pulse for preparation of the bound wave packet, such as a few-cycle UV or attosecond pulse. Previous experiments along these lines (e.g., [41] [53]) were conducted in the single-photon dressing regime and did not study the influence of the light induced potential surfaces on the angular dependence of dissociation.

Our approach allows us to follow the variation of the LIPs throughout the dressing laser field. On the timescale of the mid-IR pulse envelope we observe the opening and closing of dissociation pathways as the dressing field strength changes. On shorter time scales, the propagation of the dissociating wave packet will become accessible with sub-femtosecond time resolution by monitoring the electron localization on either fragment. More generally, we have shown how complex LIP energy landscapes determine the outcome of molecular dissociation, using $H_2$ as an example. Our approach will allow for elucidating the reaction dynamics of more complex molecules in the presence of LICIs and higher-order point intersections.

# Methods

## Experiment

The employed experimental technique is a variant of Ref. [51]. The output of a commercial Ti:Sa chirped pulse amplification (CPA) laser (Coherent Elite, 10 kHz, 2 mJ) is split into two parts. The stronger part (85%) is used to pump an optical parametric amplifier, in order to obtain carrier-envelope phase (CEP) stable idler pulses at 2.3 µm. The second part of the CPA output is focused into an argon-filled hollow core fiber to obtain broadband laser pulses, which are subsequently compressed to a pulse duration of ~5 fs. The laser pulses are recombined using a polished Si mirror (thickness 2.2 mm) at 60° angle of incidence.

After recombination, the pulses are focused in the center of a Cold Target Recoil Ion Momentum Spectrometer (COLTRIMS) [54]. The intensity of the mid-IR pulse is weak enough to not cause any notable ionization by itself. Because ions are only produced in the small focal volume of the visible pulse (1/e² width (7 ±2) µm), focal volume averaging within the larger focal volume ((30 ± 10) µm) of the mid-IR pulse is essentially avoided. In the COLTRIMS, the three-dimensional (3D) momenta of ions and electrons generated in the laser focus are measured in coincidence, which provides access to the recoil-frame ion momentum that arises solely from the nuclear dynamics on the LIPs. See Supplementary Figure 5 for a comparison of laboratory-frame and recoil-frame measurements. The measurement of the delay dependence of the electron momentum distribution yields the instantaneous mid-IR vector potential at each delay value, as shown in Supplementary Figure 4.

## Time-dependent Schrödinger equation

For the dynamics in the $H_2^+$ cation, we solve a two-dimensional (one angle and one bondlength) Schrödinger equation that includes dipole coupling between the two relevant electronic states $^2\Sigma_g^+$ and $^2\Sigma_u^+$

$$i\frac{\partial}{\partial t}\begin{bmatrix}\Psi_g(\vec{R})\\ \Psi_u(\vec{R})\end{bmatrix} = -\frac{1}{2\mu}\left(\frac{\partial^2}{\partial R^2}+\frac{1}{R^2}\frac{\partial^2}{\partial \theta^2}\right)\begin{bmatrix}\Psi_g(\vec{R})\\ \Psi_u(\vec{R})\end{bmatrix} + \begin{bmatrix}V_g(R) & -\vec{F}(t)\cdot\vec{d}(R)\\ -\vec{F}(t)\cdot\vec{d}(R) & V_u(R)\end{bmatrix}\begin{bmatrix}\Psi_g(\vec{R})\\ \Psi_u(\vec{R})\end{bmatrix}, \quad (1)$$

where $\vec{R} = (R,\theta)$ are the bondlength and angle between the laser field and the molecular axis. $\vec{F}(t)$ is the electric field of the laser that couples the two electronic states. The form of the electronic potential energy curves $V_g$ and $V_u$, as well as the transition dipole $\vec{d}$, are taken from Bunkin and Tugov [55]. Equation (1) was solved numerically using the Fourier split-operator method.

In our experiment, the $H_2^+$ system is created starting from the $H_2$ neutral through strong field ionization. The initial state of the wave function of the ionic simulations assumes a vertical transition from the ground electronic ($^1\Sigma_g$) and ground vibrational state of the $H_2$ neutral to the ground electronic state of the ion. The ground vibrational state on the $^1\Sigma_g$ of the neutral is modeled as Morse oscillator state using Morse parameters derived from Herzberg [56]. The rotational degree of freedom was initialized to a thermal rotational distribution, with temperature chosen to be low enough such that only the |J=0> rotational ground state is populated.

The laser field used in the calculations presented in Fig. 3 can be expressed as

$$E(t) = E_{IR}(t + \Delta t) + E_{VIS}(t) + E_{ped}(t)$$

where $\Delta t$ is the time delay between visible and mid-IR pulses and each pulse, E(t), is given by an expression (using atomic units),

$$E(t) = \sqrt{I}\exp\left(-2\ln 2\left(\frac{t}{\tau}\right)^2\right)\cos(\omega t + \varphi),$$

where $\varphi$ is the carrier-envelope phase (CEP).

The laser field consists of a mid-IR pulse ($\lambda_{IR}$=2300 nm, $\tau_{IR}$=60 fs, $I_{IR}$=30 TW/cm²), an ionizing few-cycle visible pulse ($\lambda_{VIS}$=730 nm, $\tau_{VIS}$=5 fs, $I_{VIS}$=300 TW/cm²), and a visible pulse pedestal ($\lambda_{ped}$=730 nm, $\tau_{ped}$=60 fs, $I_{ped}$=10 TW/cm²)). The calculations are started at t=0, i.e. in the center of the visible pulse and performed for various values of $\Delta t$ and, for each $\Delta t$, $\varphi_{VIS} = \varphi_{ped} = n\pi$ (n=0, 1) and $\varphi_{IR} \equiv 0$.

## Floquet states

For each molecular alignment angle θ, the R-dependent Floquet energy curves are calculated for a field

$$E(t,\theta) = \sqrt{I_{IR}} \cos(\omega t) \sin\theta + \sqrt{I_{ped}} \cos(3\omega t + \phi) \cos\theta,$$

where $\phi$ is the relative phase of the two fields. The potential energy landscape presented in Fig. 3(b) is for the relative phase $\phi = 0$.

At each point along R, the Floquet states were constructed as follows. First, the one-period propagator $U(t,t+\tau;R)$, where $\tau=2\pi/\omega$ is the period of the 2300 nm laser field, was constructed numerically using

$$U(t, t+\tau; R) = e^{-iH_e(R,t_{N-1})\Delta_t} e^{-iH_e(R,t_{N-2})\Delta_t} \ldots e^{-iH_e(R,t_1)\Delta_t} e^{-iH_e(R,t_0)\Delta_t},$$

where the time interval $\tau$ has been split into $N$ time steps of duration $\Delta_t = \tau/N$ with the intermediate times given by $t_n = t + n\Delta_t$, and the purely electronic Hamiltonian $H_e(t)$ is given by

$$H_e = \begin{bmatrix} V_g(R) & -\vec{F}(t)\cdot\vec{d}(R) \\ -\vec{F}(t)\cdot\vec{d}(R) & V_u(R) \end{bmatrix}.$$

The Floquet states $|F_\alpha(R,t)\rangle$ are then the eigenstates of $U(t,t+\tau;T)$,

$$U(t,t+\tau;R)|F_\alpha(R,t)\rangle = e^{-i\varepsilon_\alpha(R)t}|F_\alpha(R,t)\rangle,$$

Where the $\varepsilon_\alpha(R)$ are the quasi-energies of the Floquet states $|F_\alpha(R,t)\rangle$. The Floquet states and quasi-energies are found by directly diagonalizing the 2×2 $U(t, t+\tau; R)$ matrix for each $R\hat{}$. These quasi-energies (or Floquet energies) are what is referred to as Light-Induced Potentials in the main text.

*Acknowledgements*

The authors thank D. Crane, R. Kroeker, and B. Avery for technical assistance. We thank F. Bouakline, M. Richter, A. M. Sayler, M. F. Kling and B. Bergues for fruitful discussions. This project has received funding from the EU's Horizon2020 research and innovation programme under the Marie Sklodowska-Curie Grant Agreement No. 657544. Financial support from the National Science and Engineering Research Council Discovery Grant No. 419092-2013-RGPIN is gratefully acknowledged.

*Author contributions*

M.K., Z.D. and A.S. conceived and conducted the experiment, and analysed the results. M.S., M.K., S.C., M.J.J.V, and D.M.V. performed simulations and interpreted the data with P.B.C and A.S. All authors discussed the results and contributed to the final manuscript.

*Conflict of interest*

The authors declare no competing interests.